\documentclass[twocolumn]{webofc}
\usepackage[varg]{txfonts}
\usepackage{amsmath,amssymb}    % need for subequations

\usepackage{bm}
\usepackage{mathrsfs}
\usepackage{ulem}%line on text

\usepackage{color}
\makeatletter

\begin{document}

\title{Dilatancy of frictional granular materials under oscillatory shear with constant pressure}

\author{\firstname{Daisuke} \lastname{Ishima}\inst{1}\thanks{\email{daisuke.ishima@yukawa.kyoto-u.ac.jp}} \and
        \firstname{Hisao} \lastname{Hayakawa}\inst{1}
}

\institute{Yukawa Institute for Theoretical Physics, Kyoto University, Kyoto 606-8502, Japan
}

\abstract{
We perform numerical simulations of a two-dimensional frictional granular system under oscillatory shear confined by constant pressure.
We found that the system undergoes dilatancy as the strain increases.
We confirmed that compaction also takes place at an intermediate strain amplitude for a small mutual friction coefficient between particles.
We also found that compaction depends on the confinement pressure while dilatancy little depends on the pressure.
}

\maketitle

\section{Introduction\label{intro}}

Granular materials consist of a collection of distinct macroscopic particles such as sands and glass beads.
The evolution of the particles obeys Newton's equation with repulsive force between particles.
In reality, due to the roughness of the particles, mutual friction between particles is unavoidable in granular systems.
We call such particles frictional particles.
However, we sometimes ignore the mutual friction as an idealistic model.
We call such particles frictionless particles.
So far many theoretical studies have only been conducted for frictionless particles.
Previous studies, however, pointed out that the mutual friction causes drastic changes in rheology such as the emergence of discontinuous shear thickening (DST) \cite{Brown09, Otsuki11, Seto13, Kawasaki18}.
Bi et al. found that the shear jammed state is only observable in frictional particles \cite{Bi11}.
Various researchers studied shear jammed state and DST in systems taking into account the mutual friction \cite{Zhang08,Zhang10, Sakar13,Fall15,Sarkar16, Peters16, Wang18,Otsuki20}.

It is well known that dilatancy which is the volume expansion of a collection of particles takes place in granular systems \cite{Reynolds1885,Bagnold66,Thompson91,Jaeger96}, when shear is applied to them.
More dilatant structure is realized for the high shear rate \cite{MiDi04,Cruz05,Forterre08,Boyer11}.
Previous studies also reported the existence of  compaction in which repeated oscillations to the system make particles denser and more ordered configurations \cite{Knight95,Nowak98,Haw98,Haw98v2,Nicolas00,Pouliquen03}.
Although some studies investigated the dilatancy and compaction at a fixed friction coefficient \cite{Nowak98,Nicolas00},
it is not clear how the dilatancy and compaction depend on the friction coefficient.
Thus, this paper, which is complementary to the previous paper \cite{Ishima20}, focuses on appearances of dilatancy and compaction of frictional particles under oscillatory shear.

The contents of this paper are as follows.
In the next section, we explain the setup of our numerical simulation.
In Section 3, we present our numerical results to clarify the conditions of occurrence of dilatancy and compaction.
In the last section, we summarize our results.

\section{Set up of our simulation\label{sec2}}

We consider a two-dimensional system containing $N$ granular particles with the mutual friction characterized by Coulomb's friction coefficient $\mu$.
To prevent crystallization, the system contains four-dispersion particles whose diameters are $d_{0},\ 0.9d_{0},\ 0.8d_{0},$ and $0.7d_{0}$, respectively.
There are $N/4$ particles for each particle size \cite{Luding01,Otsuki09,Otsuki11}.
Since we assume that the density of each particle is identical, the mass of each particle is proportional to the square of its diameter.
In our study, $m_{0}$ stands for the mass of a particle with its diameter $d_0$ \cite{Ishima20}.

\begin{figure}[htbp]
  \centering
    \includegraphics[clip,width=8.5cm]{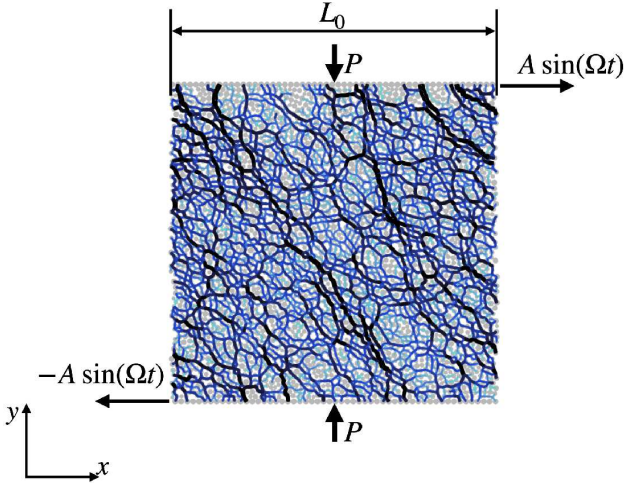}
    \caption{
    A snapshot of our simulated system, where $P$, $A$ and $\Omega$ are the pressure, strain amplitude, and angular frequency of oscillation, respectively.
	The circles and lines represent the particles and contact network of particles, respectively.
    The thickness of each line is proportional to the magnitude of the contact force.
    }
    \label{fig1}
\end{figure}

We ignore the effect of gravity throughout this paper.
We adopt the periodic boundary condition in the shear $(x-)$direction and introduce mobile walls in the vertical $(y-)$direction.
The top and bottom walls consist of particles of diameter $d_{0}$, in which the number of particles of each wall is $N_{\textrm{w}}$ and their centers of gravity are located at $x_{G}^{\pm}(t)=\pm A\sin(\Omega t)$ and $y_{G}^{\pm}(t)=\pm L_{y}(t)/2$, with the amplitude $A$ and angular frequency $\Omega$ of oscillation (see Fig.\ref{fig1}).
Since the system size $L_y(t)$ under the influence of the confinement pressure $P$ depends on time, we introduce an effective shear strain:
\begin{align}
\gamma_{0,\textrm{eff}}:=\frac{2A}{L_{0}},
\end{align}
where $L_0$ is the linear size of the shear direction.

We adopt the equations of motion of the walls at $y_{G}^{\pm}(t)$:
\begin{align}
m_{\textrm{w}}\frac{dv_{\textrm{w},y}^{\pm}}{dt}=\pm(P_{\textrm{w}}^{{\pm}} - P)L_{0}-\xi_{\textrm{d}}  v_{\textrm{w},y}^{\pm}\label{eq:walls},
\end{align}
where $\pm$, $m_{\textrm{w}}$, $v_{\textrm{w},y}^{\pm}$, $P_{\textrm{w}}^{\pm}$, and $\xi_{\textrm{d}}$ are indices referring to the wall at $y^{\pm}_{G}(t)$ , mass of the wall $m_{\textrm{w}}=N_{\textrm{w}}m_{0}$, velocity in the $y-$direction of the wall at $y^{\pm}_{G}(t)$, internal pressures acting on the walls at $y_{G}^{\pm}(t)$, and damping coefficient, respectively\footnote{ The damping coefficient $\xi_{\textrm{d}}$ is neccesary to stabilize the motion of walls.}.

Let $m_i$, $\bm{r}_i$, $ I_i=m_i d_i^2/8$, and $\omega_{i}$ be the mass, position, moment of inertia, and $z-$component of rotational velocity of particle $i$, respectively.
 The equations of motion for translation and rotation are, respectively, given by
\begin{align}
m_i \frac{d^{2}\bm{r}_i}{dt^{2}}=\sum_{j\neq i}\bm{f}_{ij}, \label{eqTra}\\
I_i\frac{d\omega_i}{dt}=\sum_{j\neq i}T_{ij}, \label{eqRot}
\end{align}
where we have introduced the contact force $\bm{f}_{ij}$ and torque $T_{ij}$ exerted by particle $j$ on particle $i$.
The torque $T_{ij}$ in Eq. \eqref{eqRot} satisfies
\begin{align}
T_{ij}=-\frac{d_i}{2}\bm{f}_{ij}\cdot\bm{t}_{ij}, \label{eqTorque}
\end{align}
where $\bm{t}_{ij}$ is the tangential vector between particles $i$ and $j$.
We adopt the Cundall-Strack model for the contact force between particles \cite{Cundall79,Luding08}, where the contact force $\bm{f}_{ij}$ is expressed as
\begin{align}
\bm{f}_{ij}=\left(\bm{f}_{ij,n}+\bm{f}_{ij,t}\right)\Theta(d_{ij}-r_{ij}) \label{eqForce}
\end{align}
with $d_{ij} = (d_i + d_j) / 2$ and $r_{ij} = |\bm{r}_i-\bm{r}_j|$.
Here we have introduced the normal force $\bm{f}_{ij,n}$, the tangential force $\bm{f}_{ij,t}$, and Heaviside's step function $\Theta(x)$ satisfying $\Theta(x)=1$ for $x\geq 0$ and $\Theta(x)=0$ otherwise.
The normal repulsive force is modeled as an elastic force with a linear spring constant $k_{n}$ and the dissipative force with a damping constant $\xi_{n}$.
Then, we model the normal force as
\begin{align}
\bm{f}_{ij,n}&=k_n\delta_{ij,n}\bm{n}_{ij}-\xi_n\bm{v}_{ij,n},\label{eq:fn}
\end{align}
where we have introduced $\bm{n}_{ij}=\bm{r}_{ij}/r_{ij}$ and $\bm{v}_{ij,n}=(\bm{v}_{ij}\cdot\bm{n}_{ij})\bm{n}_{ij}$ with $\bm{r}_{ij}=\bm{r}_{i}-\bm{r}_{j}$, $\bm{v}_{ij}=d\bm{r}_{ij}/dt$ and $\delta_{ij,n}=d_{ij}-r_{ij}$ \footnote{Note that this modeled normal force could be a negative value \cite{PoschelBook}. }.
On the other hand, the tangential force $\bm{f}_{ij,t}$ takes two different expressions depending on the magnitude of the tangential force $f_{ij,t}=|\bm{f}_{ij,t}|$:
\begin{align}
  \bm{f}_{ij,t} &= \begin{cases}
    k_t\delta_{ij,t}\bm{t}_{ij}-\xi_t\bm{c}_{ij} & ( f_{ij,t}<\mu f_{ij,n} ) ,\\
    \mu f_{ij,n}\bm{t}_{ij} & (\textrm{otherwise})\label{eqFri},
  \end{cases}
\end{align}
where $\delta_{ij,t}$ is the tangential displacement between the particles.
We have introduced the spring constant in the tangent direction $k_{t}$,
 damping coefficient in the tangent direction $\xi_t$, magnitude of the normal force $f_{ij,n}=|\bm{f}_{ij,n}|$, and tangential velocity at the contact point $\bm{c}_{ij}:=\bm{v}_{ij}-\bm{v}_{ij,n}+\bm{t}_{ij}(d_i\omega_i+d_j\omega_j)/2$.

In our simulation, we adopt $k_{t}=k_{n}/2$, $\xi_{n}=(m_{0}k_{n})^{-1/2}$, $\xi_{t}=\xi_{n}$, and $\xi_{\textrm{d}}=\xi_{n}$ \cite{Otsuki20,Cruz05}.
The control parameters of our system are $\hat P:=P/k_n$, $\gamma_{0,\textrm{eff}}$, $\Omega\sqrt{m_0/k_n}$, and $\mu$.
We deal with $\hat P$ from $2.0\times10^{-5}$ to $2.0\times10^{-3}$, $\gamma_{0,\textrm{eff}}$ from $1.0\times10^{-6}$ to $1.0$, and $\mu$ from $0$ to $1.0$.
Furthermore, we mainly investigate that $N=4,000$, and $\Omega/(2 \pi) = 1.0\times10^{- 4} \sqrt{k_{n}/m_{0}}$ in this paper.
To know $\Omega$ dependences see Ref.~\cite{Ishima20}.

To prepare the initial configuration, we place particles with diameters of $0.6d_{0},\ 0.5d_{0},\ 0.4d_{0}, \textrm{and}\ 0.3d_{0}$ at random and then increase the diameter of the particles to reach the area fraction $\phi_{\textrm{ini}}=0.82$. 
Next, the confinement pressure is applied to both walls to achieve a steady state.
Then, oscillatory shear is applied through the walls~\cite{Ishima20}.
We use the symplectic Euler method with the time step $\Delta t=0.05\sqrt{m_{0}/k_{n}}$.

We have averaged the data over $10$ cycles after $N_{\textrm{ini}}=$ $10$ cycles to remove the effect of specific initial configurations.
We set $t = 0$ at the moment of the end of $N_{\textrm{ini}}$ cycles.

Let us introduce the area fraction $\phi(\Omega t,\ \hat{P},\ \gamma_{0,\textrm{eff}})$
\begin{align}
\phi(\Omega t,\ \hat{P},\ \gamma_{0,\textrm{eff}})=\frac{ \sum_{i=1}^{N}\pi d_{i}^{2}+N_{\textrm{w}}\pi d_{0}^{2} }{4L_{x}L_{y}(\Omega t,\ \hat{P},\ \gamma_{0,\textrm{eff}})}.
\end{align}
To characterize the density change we also introduce $\delta\phi$:
\begin{align}
\delta\phi(\hat P,\ \gamma_{0,\textrm{eff}}):= &\phi(\Omega t=2n\pi,\ \hat P,\ \gamma_{0,\textrm{eff}})
-\phi_{0}(\hat P)
\label{10}
\end{align}
where $n$ and $\phi_{0}(\hat P)$ are a non-negative integer and the area fraction without shear at $\hat P$, respectively.
Since our system is confined by the constant pressure, the density depends on time.
Then, we focus on the excess fraction $\delta\phi$ at the instance of zero strain and the maximum strain rate.

\section{Results of our simulation\label{sec3}}
\begin{figure}[htbp]
  \centering
    \includegraphics[clip,width=8.5cm]{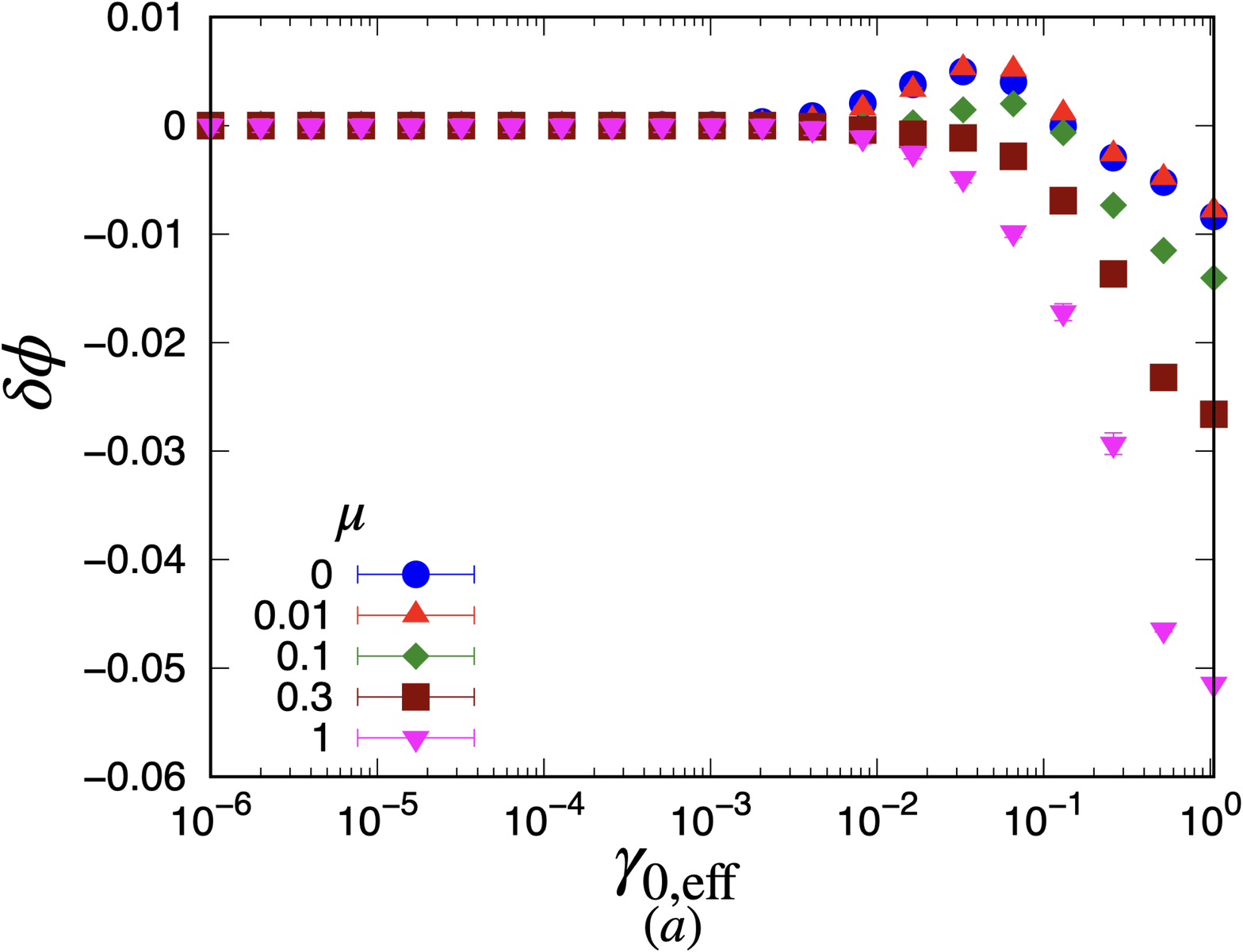}
    \includegraphics[clip,width=8.5cm]{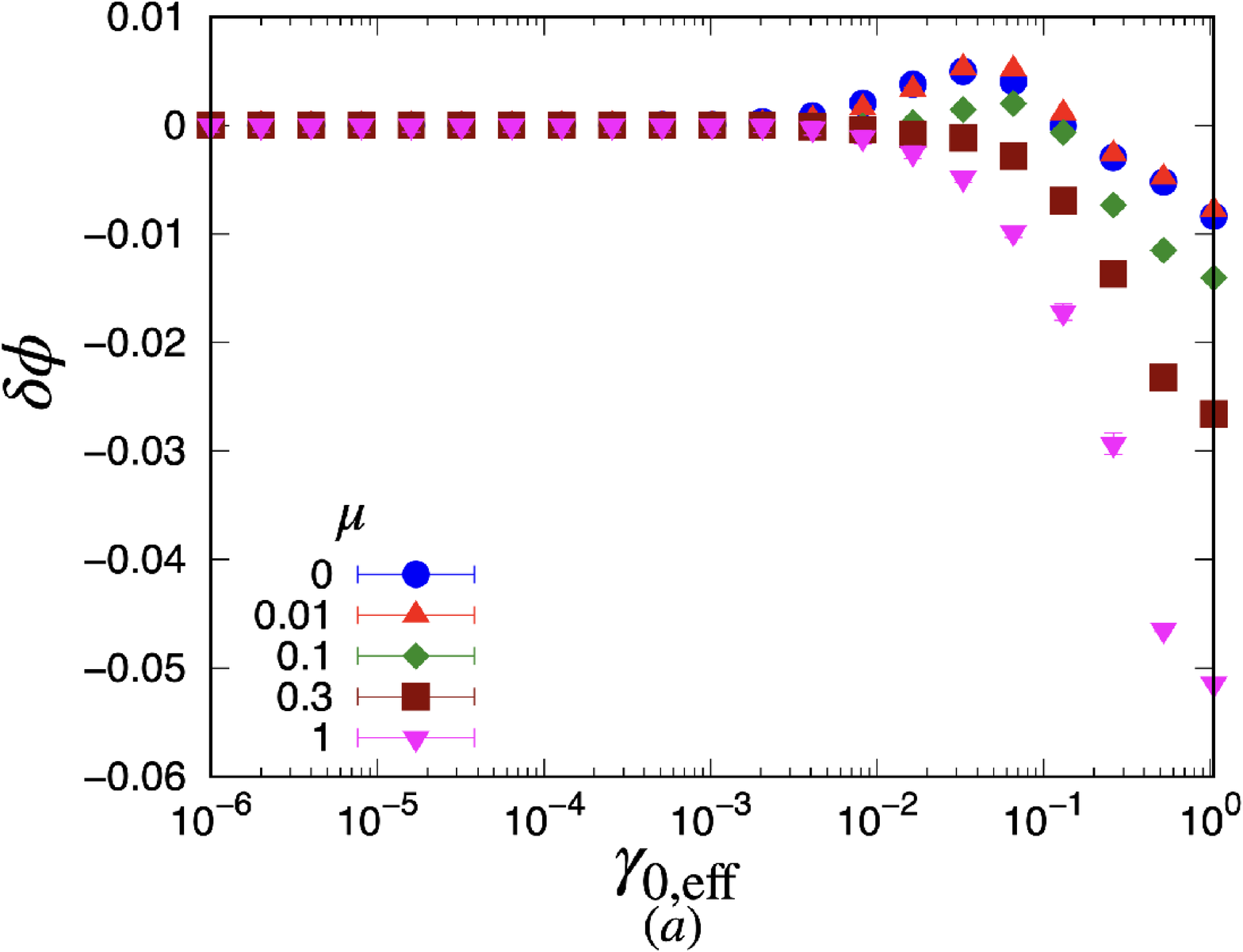}
    \caption{
    Plots of $\delta\phi$ against $\gamma_{0,\textrm{eff}}$ for various $\mu$ at (a) $\hat P=2.0\times10^{-3}$ and (b) $\hat P=2.0\times10^{-4}$.
    }
    \label{fig:phimu}
\end{figure}

Let us present the results of our simulation for various $\mu$.
We found that dilatancy is always observable for large strain amplitude (see Fig. \ref{fig:phimu}).
We also found that dilatancy is enhanced as the friction coefficient increases.
This result may be related to the fact that the region of shear jamming is more extensive for large friction coefficients \cite{Otsuki20}.
Note that the excess fraction $\delta \phi$ defined by Eq. \eqref{10} does not linearly decrease with the shear rate, as in dilatancy in steady sheared systems under constant pressures \cite{Cruz05,Forterre08}.
It is noteworthy that compaction appears at an intermediate strain amplitude for small friction coefficients.

\begin{figure}[htbp]
  \centering
    \includegraphics[clip,width=8.5cm]{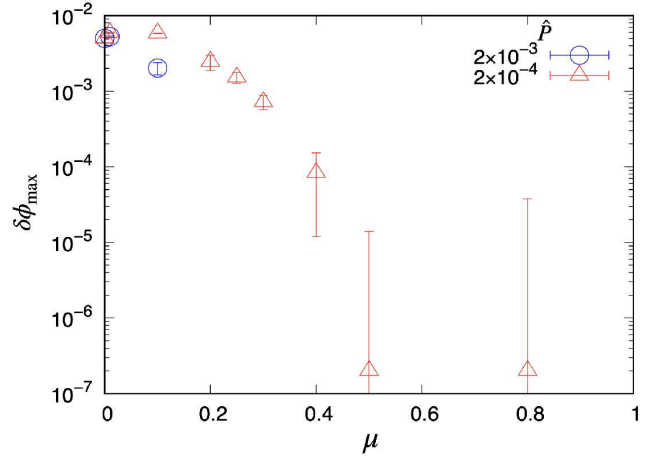}
    \caption{
    Plots of $\delta\phi_{\max}$ against $\mu$ for various $\hat P$.
    }
    \label{phiMax}
\end{figure}

To clarify the dependence of compaction on $\hat P$ and $\mu$,
we plot $\delta\phi_{\max}$, the maximum value of $\delta\phi$ at fixed $\hat P$, against $\mu$ for two different $\hat{P}$ in Fig. \ref{phiMax}.
We confirm that $\delta\phi_{\max}$ is almost independent of $\hat P$ for $\mu\leq0.01$
but it strongly depends on the pressure for $\mu\ge 0.01$.
Moreover, compaction disappears for $\mu>0.1$ at $\hat{P}=2.0\times 10^{-3}$ while it survives even for $\mu=0.4$ at $\hat{P}=2.0\times 10^{-4}$.

\begin{figure}[htbp]
  \centering
    \includegraphics[clip,width=8.5cm]{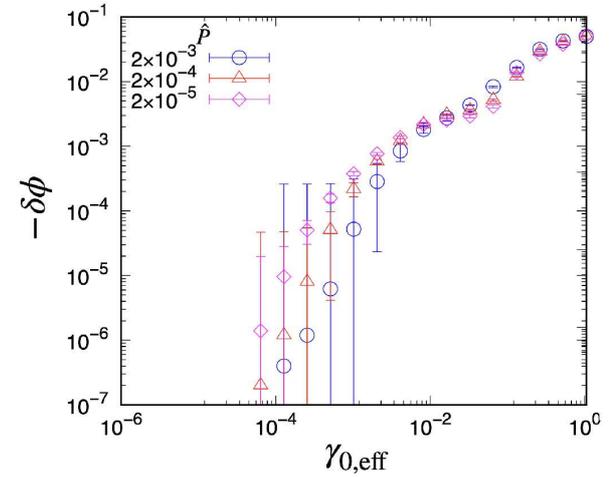}
    \caption{
    Plots of $\delta\phi$ against $\gamma_{0,\textrm{eff}}$ for various $\hat P$ at $\mu=1.0$.
    }
    \label{fig:phi}
\end{figure}
From now on, let us focus on the results for $\mu=1.0$ to discuss the pressure dependence.
From Fig. \ref{fig:phi}, $\delta \phi$ is zero for $\gamma_{0,\textrm{eff}}<10^{-4}$,
while it takes finite value for $\gamma_{0, \textrm{eff}}>10^{-4}$.
It is remarkable that $\delta\phi$ seems to depend on the pressure only a little.
In the previous paper \cite{Ishima20},
we reported that the softening point characterized by the critical strain amplitude
 $\gamma_{\textrm{S}}$ corresponding to the bending point of shear modulus can be scaled by pressure,
while the critical strain for dilatancy little depends on the pressure.
Therefore, our result suggests that there is no direct connection between the density change and the behavior of shear modulus.
Note that the trajectories of particles are reversible for the strain amplitude a little larger than $\gamma_{\textrm{S}}$ as in Refs. \cite{Bohy17,Otsuki21}.
Our results suggest that the critical condition of dilatancy is not related to the softening point but the yielding point.

\section{Concluding remarks \label{sec4}}
In conclusion, we conducted a numerical study of frictional granular systems under oscillatory shear confined by constant pressure.
We found that compaction can be observed only for small $\mu$ at an intermediate strain amplitude, while dilatancy always takes place at large strain amplitude.
We confirmed that dilatancy is a continuous nonequilibrium phase transition and little depends on the confinement pressure.

One of the most important findings in this paper is that the density changes such as dilatancy and compaction are not directly connected with the softening.
It will be important to confirm the validity of this picture in the near future. 
As another important point, our preliminary study suggests that the results strongly depend on the initial packing fraction $\phi_{\textrm{ini}}$.
Detailed studies on the $\phi_{\textrm{ini}}$-dependence will be reported elsewhere.

\vspace*{0.4cm}
\noindent

{\bf Acknowledgement}

The authors thank M. Otsuki and K. Saitoh for fruitful discussions and useful comments.
This work is partially supported by the Grant-in-Aid of MEXT for Scientific Research (Grant No. 16H04025) and the Programs YITP-T-18-03 and YITP-W-18-17.
The work of D.I. is partially supported by the Grant-in-Aid for Japan Society for Promotion of Science JSPS Research Fellow (Grant No.20J20292).
The research of H.H. has been partially supported by ISHIZUE  2020  of  Kyoto University Research Development Program.

\end{document}